# Microwave Magneto-Chiral Effect in a Noncentro-symmetric Magnet $CuB_2O_4$

Y. Nii,[*] R. Sasaki, Y. Iguchi, and Y. Onose

*Department of Basic Science, University of Tokyo 153-8902, Japan*

We have investigated microwave nonreciprocity in a noncentro-symmetric magnet $CuB_2O_4$. We simultaneously observed differently originated nonreciprocities; the classical magnetic dipolar effect and the magneto-chiral (MCh) effect. By rotating magnetic field in a tetragonal plane, we clearly unveil qualitative difference between them. The MCh effect signal reveals chiral transitions from one enantiomer to the other via intermediate achiral state. We show magnetoelectric effect plays an essential role for the emergence of microwave MCh effect.

In some cases, the velocity and decay rate of electromagnetic wave changes upon the reversal of the wave vector. In particular, in the microwave region, such directional anisotropy or nonreproicity has been utilized for microwave components such as isolators and circulators. It typically originates from a magnetic dipolar interaction in a ferromagnetic medium [1]. The asymmetric geometry of a ferromagnetic component in microwave circuits gives rise to the nonreciprocal transmittance, being irrelevant to the crystal symmetry. Recently, essentially different types of microwave nonreciprocity, which are referred to as microwave magnetoelectric (ME) effect and microwave magneto-chiral (MCh) effect, have been discerned in a chiral magnet $Cu_2OSeO_3$ [2,3] and an artificial chiral magnet [4]. These effects stem from the ME effect characteristic of systems with broken time reversal symmetry (TRS) and spatial inversion symmetry (SIS).

The directional nonreciprocity of light in the medium with broken TRS and SIS has been extensively investigated in the last two decades. Rikken and co-workers first reported the directional dichroism of visible light in a chiral molecule under magnetic field [5]. Similar directional nonreciprocities originating from material symmetry breaking have been reported in x-ray [6] and terahertz regions [7]. These optical nonreciprocities can be categorized into the optical ME effect and the MCh effect. The former arises in multiferroic materials where magnetization $M$ and electric polarization $P$ coexist. Transmissions of lights propagating parallel and antiparallel to the quantity $P \times M$ are inequivalent in this case[8, 9]. The MCh effect is induced by the interplay between crystal chirality and magnetism [5, 10-12]. The optical response (or precisely dielectric constant) varies as a function of $\gamma H \cdot k$ [13]. The $\gamma$ symbolizes chirality, and $H$ and $k$ indicate a magnetic field and wave vector of light, respectively. The microwave ME and MCh effects are the extension of these optical nonreciprocities. Considering the importance of nonreciprocity in microwave region, it is imperative to clarify how the ME and MCh driven nonreciprocities coexist with and is discriminated from the classical magnetic dipolar driven nonreciprocity in this region. Here, we report microwave MCh effect in a noncentro-symmetric magnet $CuB_2O_4$, which simultaneously breaks TRS and SIS. We successfully unveil qualitative difference between the nonreciprocity driven by MCh effect and that driven by classical magnetic dipolar effect with use of magnetic field rotation.

$CuB_2O_4$ has been known to host multiferroic property [14, 15] and to exhibit prominent nonreciprocal responses [16-19] such as giant optical ME [17, 18] and optical MCh effects [19] in a near-infrared light regime. Although the magnetic resonance in microwave regime was also reported [20, 21], the nonreciprocity has never been explored. The crystal structure has a tetragonal space group of $I\bar{4}2d$, which belongs to $D_{2d}$ point group symmetry. As shown in Fig. 1(a), a tetragonal unit cell includes two inequivalent $Cu^{2+}$ sites as denoted by Cu(A) and Cu(B). It exhibits two successive magnetic phase transitions at $T_N$ = 21 K and at $T^*$ = 9 K [22]. The first transition triggers easy-plane type Néel order at Cu(A) site. The Dzyaloshinskii-Moriya interaction induces a weak ferromagnetic component normal to the tetragonal axis. The Cu(B) moment, on the other hand, remains disorder or has only a small magnetic moment less than one fourth of that at A-site [22]. The second transition at 9 K corresponds to imcommensurate helical order both at Cu(A) and Cu(B) sites.

Our microwave measurements were mostly performed at 10 K in the weak ferromagnetic (WFM) phase, in which the canted magnetic moments induce the net in-plane magnetic moment as shown in Fig. 1 (a). When the magnetic field ($H$) rotates, all the magnetic moments are also rotated so that the net moment is along the magnetic field direction. An important feature in antiferromagnets belonging to $D_{2d}$ point group such as $CuB_2O_4$, $Ba_2CoGe_2O_7$ [23, 24] and $Ca_2CoSi_2O_7$ [25] is that the magnetic symmetry shows a large variation under rotating $H$. When $H$ points along [100] or [010] axes, the diagonal mirror symmetries in $D_{2d}$ point group are broken and chirality shows up. These $H \parallel$ [100] and $H \parallel$ [010] states have different chirality $\gamma$, which is unchanged by the 180 ° rotation of $H$. On the other hand, for $H \parallel$ [110] and $H \parallel$ [1−10], the positive and negative polarization along $z$-direction $P_z$ emerge, respectively. These symmetry variations under rotating magnetic field are not affected by the staggered component, and equivalent to that of a compressed tetrahedron having a spontaneous magnetization (The compressed tetrahedron is a motif of $D_{2d}$ crystal structure). Thus, by changing $M$ direction relative to the crystal, one can manipulate two chiral states ($\pm \gamma$) and two polar states ($\pm P_z$).

A single crystal of $CuB_2O_4$ was grown by a flux method [26], and X-ray Laue photographs were used to determine the crystallographic orientation. The sample with approximate dimensions of $3 \times 4 \times 5$ mm$^3$ was mounted on a coplanar waveguide (CPW) as shown in Fig. 1(b). Tetragonal axis is normal to the substrate of the CPW, and $a$ or $b$ axes are parallel to the microwave propagation vector $k$. Microwave transmission spectroscopy was carried out by using a vector network analyzer (Agilent E5071C), and the complex transmittance (*i.e.*, S parameters) were obtained. The transmittances of microwave propagating along +$k$ or –$k$ were denoted as $S_{21}$ (port 1 to port 2) or $S_{12}$ (port 2 to port 1), respectively. Magnetic resonance of $CuB_2O_4$ affects both amplitude and phase of the microwave, which can be detected as the change in $S$ parameters. Using a spectrum at 1000 mT as a reference at which magnetic resonance is far above our measurement range (up to 20GHz), we obtained a relative change in amplitude as $\Delta S_{ij}(B) = |S_{ij}(B)| − |S_{ij}(1000 \text{ mT})|$ and phase as $\Delta \theta_{ij}(B) = \arg[S_{ij}(B)] − \arg[S_{ij}(1000 \text{ mT})]$ (The procedure is similar to previous reports [2, 3]). So as to feed intense electromagnetic fields into the sample and to achieve enough signal-to-noise ratio, we used a signal line of 0.2 mm which is relatively narrow compared with the sample width. Thus, we assumed microwave that consists of two linearly polarized light; $H^\omega \parallel y$, $E^\omega \parallel z$ and $H^\omega \parallel z$, $E^\omega \parallel y$ [see Fig. 1(c)]. External DC magnetic fields were applied in the CPW substrate.

Figures 2(a) and 2(b) show amplitude and phase of transmission spectra in a weak ferromagnetic phase (10K) under Faraday geometry ($k \parallel \pm M$). External magnetic field of 100 mT is applied along $a$ axis, and thus $\gamma > 0$ chiral state is expected. Corresponding to magnetic resonances around 5 GHz, dip and dispersive spectra appear in amplitude and phase, respectively. As shown in the Supplemental Material [27], paramagnetic resonance at Cu(B) site would be the origin of microwave response. Notably, there are small but finite differences between $+k$ ($S_{21}$) and $-k$ ($S_{12}$) signals as shown in the insets. The difference in amplitude and phase are plotted in Figs. 2(c) and 2(d), respectively. One can clearly confirm nonreciprocal directional dichroism (NDD) and nonreciprocal directional birefringence (NDB).

In Figs. 3, we demonstrates a switching of NDD and NDB signals by systematic reversals of chirality and magnetic field. Microwave transmission measurements were performed at 10 K under Faraday geometry ($k \parallel \pm M$). So as to switch chirality, the sample was rotated 90 ° about $c$ axis as displayed in the left panel of Fig. 3. As represented in Fig. 3, one can clearly see that the sign of NDD and NDB signals were inverted by reversing either $\gamma$ or $H$. [See for example Fig. 3(a) and 3(c) for $H$ reversal, and Fig. 3(a) and 3(e) for $\gamma$ reversal.] On the other hand, when $\gamma$ and $H$ are simultaneously reversed, the polarity of nonreciprocity remains invariant. These observations follow the symmetry requirement of MCh effect as $\gamma \mathbf{H} \cdot \mathbf{k}$. This unambiguously shows that the nonreciprocity is induced by the microwave MCh effect and also demonstrates switching between two enantiomers (mirror images). Incidentally, it can be attributed to neither natural optical activity nor magneto optical activity since they must change sign under simultaneous reversal of $\gamma$ and $H$.

Figures 4(a) and 4(b) exhibit NDD spectra at various magnetic field angles for two configurations ($k \parallel a$ and $k \parallel b$). Here we simultaneously observed two different NDD, which originate from MCh effect and a classical magnetic dipolar effect. A magnetic field of 100 mT is rotated with an angle $\phi$ in the tetragonal plane. At 0°, MCh effect originated NDD spectra has positive and negative polarities for $k \parallel a$ and $k \parallel b$ configurations, respectively. At a small tilting of $|\phi| = 15$ °, one can still recognize that polarity of dominant NDD is roughly positive for $k \parallel a$ configurations while negative for $k \parallel b$ configurations. At finite $|\phi|$, however, discrepancy becomes apparent for positive and negative tilting angles. Especially, spectra are almost inverted between that at $+\phi$ and $-\phi$, when $\phi = 45$ ° or 90 °. Since MCh signal is proportional to $\gamma \mathbf{H} \cdot \mathbf{k} \propto \gamma H K \cos \phi$, the nonreciprocity is expected to be an even function of $\phi$. On the other hand, the magnetic dipolar induced NDD [1], which is common to any magnets including centrosymmetric $Y_3Fe_5O_{12}$, is expected to show different angle dependence. In the CPW, there is a mirror symmetry for the vertical plane along the center line of wave guide (*i.e.*, *x-z* plane). This mirror symmetry ensures reciprocity at $\phi = 0$ ° when the sample also has the mirror symmetry. When the magnetic field is tilted from $\phi = 0$ °, some nonreciprocity may be induced, which is reversed by the reversal of tilting direction. Thus, the MCh effect induced NDD and magnetic dipolar induced one show even and odd functions of $\phi$, respectively. In order to distinguish them, we introduce symmetric and antisymmetric NDD spectra as follows,

$$\Delta S_{sym}(\phi) = [S_{NDD}(\phi) + S_{NDD}(-\phi)]/2$$
$$\Delta S_{asym}(\phi) = [S_{NDD}(\phi) - S_{NDD}(-\phi)]/2 \quad (1)$$

where $S_{NDD}(\phi)$ spectrum is defined as $\Delta S_{21} - \Delta S_{12}$ at $\phi$. The MCh effect driven NDD should appear in $\Delta S_{sym}$. Figures 4(c) and 4(d) show summary of angular dependence of $\Delta S_{sym}$ and $\Delta S_{asym}$ evaluated from peak amplitude of spectra (See details of $\Delta S_{sym}$ and $\Delta S_{asym}$ spectra for Supplemental Material [27]). The $\Delta S_{asym}$ does not show the large difference between $k \parallel a$ and $k \parallel b$ configurations. It increases linearly with $\phi$ in the low $\phi$ region and saturates around 45°. On the other hand, the $\Delta S_{sym}$ shows characteristic behavior which seems to be proportional to $\cos 2\phi \cos \phi$ and has three nodes at $\phi = 45°$, 90°, and 135°. The signs of $\Delta S_{sym}$ are different between $k \parallel a$ and $k \parallel b$ configurations.

As reported previously [24, 25, 28, 29], the NDD and NDB can emerge when the linear ME effect is included in Maxwell's equations. In our set-up, existence of $\chi^{me}_{yz} + \chi^{em}_{zy}$ and $\chi^{me}_{zy} + \chi^{em}_{yz}$ tensors gives rise to the finite difference of complex refractive indices for counter propagating microwave. Here $\chi^{me}_{ij}$ ($\chi^{em}_{ij}$) are the $ij$ components of magnetoelectric (electromagnetic) tensors at microwave frequency. Magnetic symmetry arguments can show that these ME tensors become finite for $\boldsymbol{M} \parallel [100]$, while vanish for $\boldsymbol{M} \parallel [010]$. This ensures the existence of MCh effect at $\phi = 0°$ and disappearance at $\phi = 90°$. In addition, this is consistent with a phenomenological relation of $\boldsymbol{H} \cdot \boldsymbol{k} \propto \cos \phi$. The nodes at $\phi = 45°$ and 135°, on the other hand, directly reflect the reversal of handedness from $\gamma > 0$ to $\gamma < 0$ (or vice versa) through intermediate $\gamma = 0$ state. Strictly speaking, for $\boldsymbol{M} \parallel [110]$ and $[1-10]$, the ME tensors $\chi^{me}_{zy} + \chi^{em}_{yz}$ does not vanish and thus NDD is not symmetrically forbidden for one of the linearly polarized microwave ($H^\omega \parallel z$, $E^\omega \parallel y$). This might be reason of small finite value of $\Delta S_{sym}$ at $\phi = 45°$ and 135°.

To summarize, we have revealed directional dichroism and birefringence in microwave regime in a noncentro-symmetric magnet $CuB_2O_4$. We have simultaneously observed two different nonreciprocities, which can be ascribed to MCh effect and a classical magnetic dipolar effect. By using the variation of magnetic point group symmetries under rotating magnetic field, we have successfully separated two contributions and extracted MCh signal characteristic of $D_{2d}$ antiferromagnet. Our demonstration shows a new perspective about the classical dipolar-type and emergent spin-orbit type microwave nonreciprocities in media without TRS and SIS.

This work is partly supported by Grant-in-Aids for Young Scientists B (No. 15K21622) and for Scientific Researches (A) and (B) (No. 25247058 and No. 16H04008), and Yamada Science Foundation. Y.I. is supported by JSPS fellows (No. 16J10076).

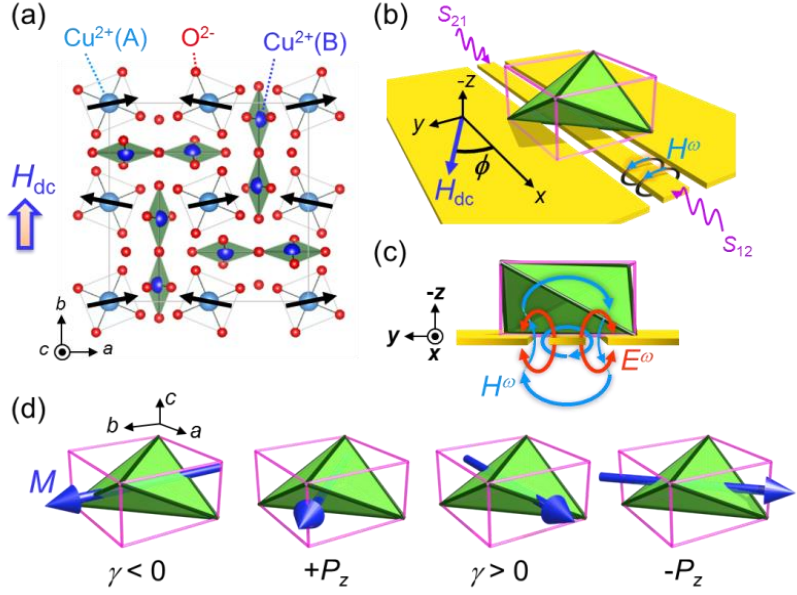

FIG. 1 (color online) (a) Crystal and magnetic structure of $CuB_2O_4$ in a weak ferromagnetic phase under $H \parallel b$. (b) Experimental configuration of microwave transmission spectroscopy. Compressed tetrahedron (green) is a motif of $CuB_2O_4$ as a simplest model having $D_{2d}$ symmetry. (c) Electromagnetic field distribution around CPW. (d) In-plane magnetization induced chirality ($\gamma$) or polarity ($P_z$) in $CuB_2O_4$. From left to right, $M$ points along [010], [110], [100], and [1−10] directions, and magnetic point group changes from 2'22', mm'2', 22'2', to m'm2', respectively. The $\gamma > 0$ and $\gamma < 0$ states correspond to two enantiomers (mirror images) with right- and left-handedness, respectively.

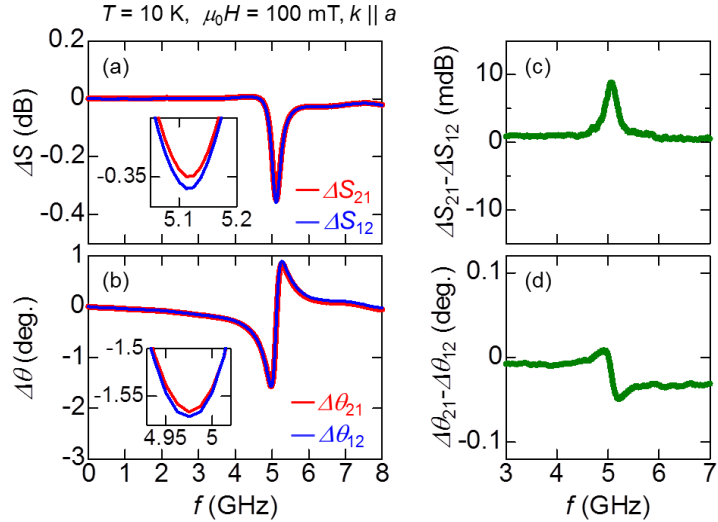

FIG. 2 (color online) Microwave transmission spectra of (a) amplitude and (b) phase at 10 K under Faraday geometry. Differences between counter-propagating microwaves shown in red and blue lines manifests nonreciprocity. (c) NDD and (d) NDB spectra derived from (a) and (b), respectively.

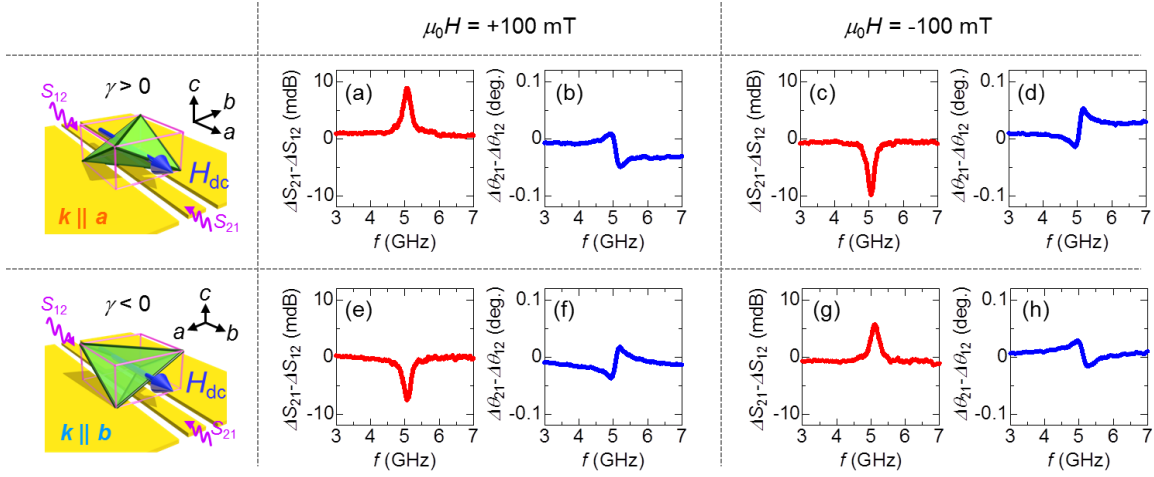

FIG. 3 (color online) (a)-(h) Switching of nonreciprocal signal by reversals of chirality and magnetic fields. Chirality was reversed via 90 ° rotation of sample about *c* axis with keeping magnetic field direction (*H* || *k*) as illustrated in the left-top and the left-bottom. These configurations are denoted as *k* || *a* and *k* || *b*. NDD and NDB signals systematically displayed in four different combinations of chirality and magnetic fields: $\gamma > 0$, $H > 0$ [(a) and (b)], $\gamma > 0$, $H < 0$ [(c) and (d)], $\gamma > 0$, $H > 0$ [(e) and (f)], and $\gamma > 0$, $H > 0$ [(g) and (h)]. They show sign change by reversing either *H* or $\gamma$, while are invariant under simultaneous reversal of them.

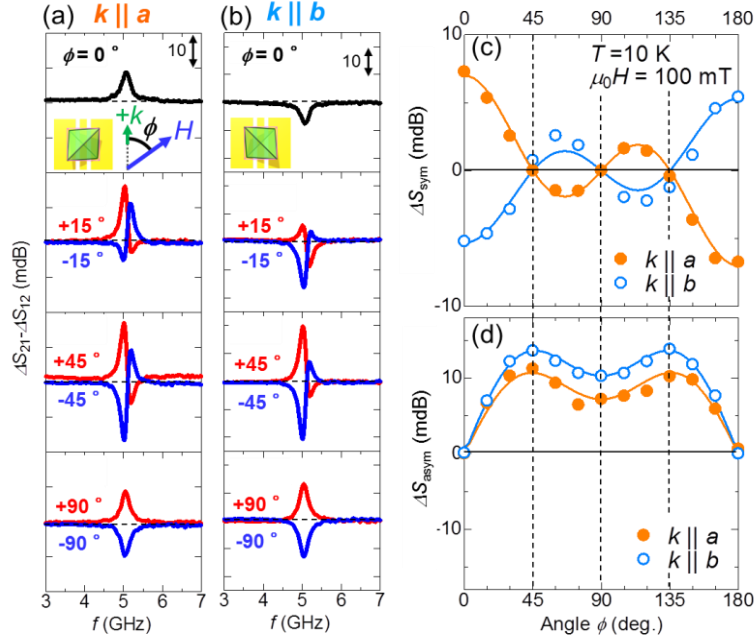

FIG. 4 (color online) (a), (b) Angular dependence of NDD spectra at 10 K under 100 mT for (a) *k* || *a* and (b) *k* || *b* configurations. (c), (d) Angular dependence of (c) symmetric and (d) antisymmetric components of NDD amplitude, respectively. The symmetric signal can be ascribed to MCh effect, while the antisymmetric one the magnetic dipolar-type nonreciprocity (see text). Nodes of $\Delta S_{\text{sym}}$ at $\phi = 45°$ and $135°$ indicate sign changes of $\gamma$ across achiral ($\gamma = 0$) state. Lines are phenomenological $\cos 2\phi \cos \phi$ dependence.

# Supplemental Material for Microwave Magneto-Chiral Effect in a Noncentro-symmetric Magnet $CuB_2O_4$

Y. Nii,[*] R. Sasaki, Y. Iguchi, and Y. Onose

*Department of Basic Science, University of Tokyo 153-8902, Japan*

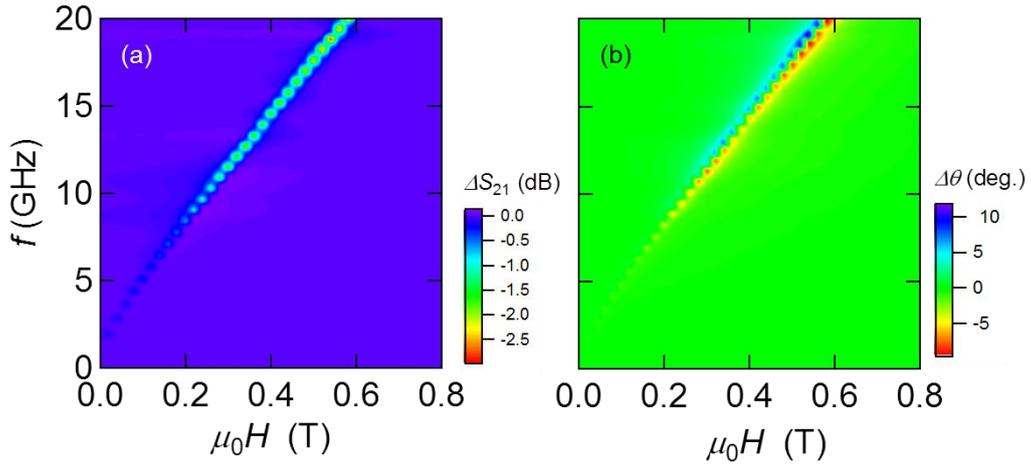

FIG. S1. (a), (b) Color mappings of (a) $\Delta S_{21}$ and (b) $\Delta \theta_{21}$ spectra as a function of a magnetic field at 10 K.

Figures S1(a) and S1(b) show color maps of amplitude $\Delta S_{21}(B)$ and phase $\Delta \theta_{21}(B)$ spectra in *f-H* plane. The temperature is 10 K and magnetic field is applied along *a* axis. The amplitude and phase spectra at 1 T can be utilized as background spectra. These figures illustrate the linear magnetic field dependence of resonant frequency.

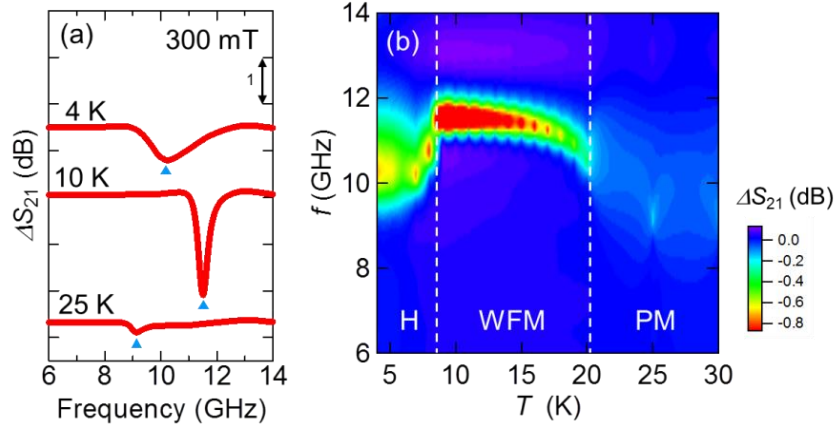

FIG. S2. Temperature dependence of $\Delta S_{21}$ with a magnetic field of 300 mT. (a) $\Delta S_{21}$ spectra of helical (4 K), weak ferromagnetic (10 K), and paramagnetic (25 K) phases. Offsets are added to these spectra for clarity. (b) Color plot of $\Delta S_{21}$ spectra versus temperature at 300 mT. H, WFM, and PM denote the helical, the weak ferromagnetic, and the paramagnetic phases, respectively.

Figure S2(a) shows typical transmission spectra in the helical (4K), the weak ferromagnetic (10K), and paramagnetic (25 K) phases, respectively. Figures S2(b) exhibit the temperature dependence of $\Delta S_{21}$ at 300 mT. Magnetic resonance can be seen in helical and WFM phases and even in a disordered PM phase. We performed detailed nonreciprocity measurement in the WFM phase since the largest and the sharpest dip arise in the WFM phase.

From the linear magnetic field dependence with small intercept and the persistence in paramagnetic phase, we assign the origin of observed microwave response to electron paramagnetic resonance of $CuB_2O_4$. In the weak ferromagnetic phase at 10 K, Cu(A) spins have canted antiferromagnetic order whereas Cu(B) spins remains disorder or has only small magnetic moment [1]. Although further investigation might be still needed to make a final conclusion, it can be attributed to electron paramagnetic resonance at Cu(B) sites. It should be noted that the assignment of origin of magnetic resonance does not affect our discussion about the nonreciprocity based on the magnetic symmetry.

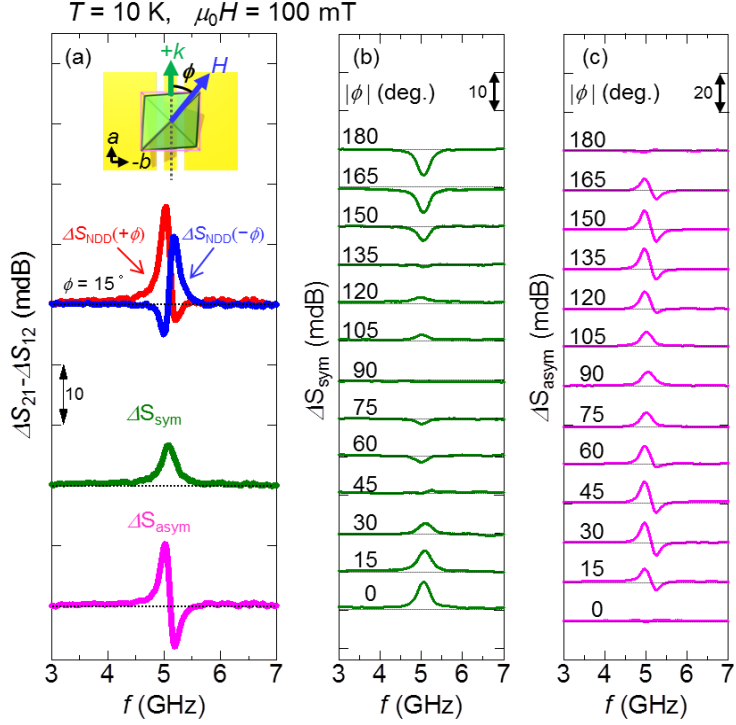

FIG. S3. Analysis of NDD spectra under tilted magnetic fields of 100 mT at 10 K. Propagation vector is parallel to the *a* axis. (a) NDD spectra at $\phi = \pm 15°$. Using two data denoted as $S_{NDD}(+\phi)$ and $S_{NDD}(-\phi)$, one can extract symmetric ($\Delta S_{\mathrm{sym}}$) and antisymmetric ($\Delta S_{\mathrm{asym}}$) components about $\phi$. The former is derived by $\{S_{NDD}(+\phi) + S_{NDD}(-\phi)\}/2$, while the latter is derived by $\{S_{NDD}(+\phi) - S_{NDD}(-\phi)\}/2$. (b), (c) Symmetric and antisymmetric components of NDD spectra at various $|\phi|$. Polarity of $\Delta S_{\mathrm{sym}}$ peaks change sign at 45°, 90° and 135°.

Figures S3 show angular dependence of NDD spectra at 10 K. Magnetic field is tilted by an angle of $\phi$ in the tetragonal plane. In order to separate MCh effect originated spectra from classical dipolar originated ones, we measure spectra for positive and negative tilted angle of the field ($+\phi$ and $-\phi$), which are denoted as $S_{NDD}(+\phi)$ and $S_{NDD}(-\phi)$, respectively. Then we obtained $\Delta S_{\mathrm{sym}} = \{S_{NDD}(+\phi) + S_{NDD}(-\phi)\}/2$ and $\Delta S_{\mathrm{asym}} = \{S_{NDD}(+\phi) - S_{NDD}(-\phi)\}/2$. Note that MCh effect follows an even function of $\phi$, whereas magnetic dipolar nonreciprocity should have an odd function of $\phi$. With use of these procedures for every spectrum at $\pm\phi$, we have derived a series of $\Delta S_{\mathrm{sym}}$ and $\Delta S_{\mathrm{asym}}$ spectra as shown in Figs. S3(b) and S3(c). Importantly, sign change of $\Delta S_{\mathrm{sym}}$ appears at 45°, 90° and 135°, which is characteristic of $D_{2d}$ symmetry.